# Inequality in Educational Attainment: Urban-Rural Comparison in the Indian Context

Sangita Das

Department of Economics, Syamaprasad College

Kolkata-700026, West Bengal, India

ORCID link: https://orcid.org/0000-0003-0109-1955

E-mail address: dassangita844@gmail.com or sangitadaseco@gmail.com

Phone Number: (+91) 9830906540

# Inequality in Educational Attainment: Urban-Rural Comparison in the Indian Context

**Abstract:**

The article tries to compare urban and rural literacy of fifteen selected Indian states during 1981 - 2011 and explores the instruments which can reduce the disparity in urban and rural educational attainment. The study constructs the Sopher's urban-rural differential literacy index to analyze the trends of literacy disparity across fifteen states in India over time. Although literacy disparity has decreased over time, Sopher's index shows that the states of Andhra Pradesh, Madhya Pradesh, Gujarat, Odisha, Maharashtra and even Karnataka faced high inequality in education between urban and rural India in 2011. Additionally, the Fixed Effect panel data regression technique has been applied in the study to identify the factors which influence urban-rural inequality in education. The model shows that the following factors can reduce literacy disparity between urban and rural areas of India: low fertility rate in rural women, higher percentages of rural females marrying after the age of 21 years, mother's educational attainment and their labour force participation rate in rural areas.

## 1. Introduction

Education is the most powerful tool for social change, and its equitable distribution between urban and rural population has positive implications for the country's social welfare (Dreze & Sen, 1999). Literacy rate is usually measured by taking the number of adults who are literate (i.e. able to read, write and understand a language) as a percentage of the total number of adults in a country (UNESCO, 1946). It includes primary education, desirable knowledge about civic needs, personal hygiene, the adoption of political and occupational skills and the potential to participate fully in their community (UNESCO, 2004). Sen (1995) has shown that literacy is an important socio-economic indicator and it has instrumental as well as intrinsic significance for designing the country's strategy of development. Moreover, Education is valuable to a person to attain freedom in at least five distinct ways[1]-intrinsic importance, instrumental personal roles, instrumental process roles, empowerment and distributive roles (Dreze & Sen, 1999).

Even though education has an enormous importance in a country's development, India is still characterized by a low level of literacy and a high disparity in educational attainment between urban and rural areas. The United Nations "Education for All - Global Monitoring Report (2010)" declared India as the country with the highest number of illiterate adults. In the aggregate scenario, India is yet to reach 75 per cent literacy (Census of India, 2011). To promote literacy among unschooled persons in India, several educational policies such as the National Adult Education Programme (1978), New Education Policy (1986), Rural Functional Literacy Programme (1986), National Literacy Mission (1988), Post Literacy Campaign (1988), Total Literacy Campaign (1989), Right of Children to Free and Compulsory Education Act (2009) etc. have already been initiated. Despite these policy measures, the country remains home to 318 million illiterate rural people[2] (Socio-Economic and Caste Census, 2011). School

---

a) [1]Intrinsic Importance: Being educated, a person gets an opportunity of his/her effective freedom.
b) Instrumental Personal Roles: A person's education can help to find a job, and more generally, person gets human freedom, capability and an enhanced quality of life attained through the economic opportunities.
c) Instrumental Social Roles: The higher level of literacy and the basic level of education can facilitate public discussion of social needs and promote informed collective demands (e.g. health, peace and social security).
d) Instrumental Process Roles: The process of schooling can benefit almost any society. For example, the incidence of child marriage is intimately connected with the lack of school education for girls, and the expansion of schooling can reduce the distressing phenomenon of child marriage that prevalent in India.
e) Empowerment and Distributive Roles: Greater literacy and educational attainment of disadvantaged groups can increase their ability to combat dishonest actions, organize politically, and get a fairer deal.

[2] The 75th round NSSO survey on Household Social and Consumption (July 2017-June 2018) also observed that around 27 per cent rural people are still illiterate in India. Among the persons aged 3 to 35 years in rural India,

infrastructure is not equally distributed between rural and urban areas. Around 18 per cent of teachers in rural India do not have minimum qualifications for teaching (District Information System for Education data [DISE], 2016-17) and only 68 per cent of toilets in government-run rural schools are usable (Annual Status of Education Report [ASER], 2017). Apart from poor sanitation facilities, students in rural India suffer from the inaccessibility of modern tools for learning. Sometimes they cannot even attend school because of the large distance between 'home' and 'school' (Agrawal, 2014). As per the Census of India (2011), 8.4 crore children in the age group of 5-17 years do not pursue their school education and 3.81 crore children between 6-13 years do not continue school (including students who never attended and attended before).

In this context, the paper tries to analyse the trends in urban-rural differential in literacy from 1981 to 2011 and the underlying socio-economic factors to come out with policy advice for reducing this differential across the fifteen Indian states.

## 2. Relevant literature, research gap and scope of the study

Rural India has experienced economic stagnation resulting from an unequal distribution of education between urban and rural communities. This can have grave implications for society in the form of poverty, the prevalence of caste-based discrimination, and a lack of awareness (Desai & Kulkarni, 2008). Krishan and Shaym (1978) examined the regional aspects of the urban-rural differential in literacy in India and use the 1971 census data for all the 356 districts in India. The study shows that education and innovation originate in urban areas and trickle down to villages. Zhang (2006) used the empirical data from 14 school systems in sub-Saharan African countries between 2000 and 2002 and developed the reading literacy scores (students' performance on a paper-and-pencil test and their ability to understand) to measure urban-rural educational attainment in primary school. The study emphasizes that endowing rural schools with resources (more quality in physical and human resources) and parents encouraging the academic work of children are prime factors in reducing urban-rural disparities in students' learning outcomes. Jhariya and Jain (2014) analyzed the regional inequality in educational attainment in different parts of Madhya Pradesh from 2001 to 2011. They observed that other than financial assistance for weaker sections of society, women empowerment and village empowerment are essential for reducing the wide literacy gap between male-female and urban-rural population. Applying the difference-in-difference method in the Chindwara district in

---

15.7 per cent never enrolled in school, and 40.7 per cent enrolled but presently discontinued the class.

Madhya Pradesh (MP) in 2004, Afridi (2011) observed that free meal services in the public primary schools raise the school enrolment rate, especially among the girls and increase the attendance rate in rural central India. Using Heckman's two-step method, Agrawal (2012) illustrated that a higher ability to work and return from work is achieved by attending school, and it differs in the urban and rural population. Investments in schooling are dictated by assessments of future rates of return to schooling. Poor parents prefer to send their children to school if the expected return from schooling exceeds their opportunity cost of education (Becker, 2009).

Several researchers explained the possible socio-economic factors impacting urban-rural differential in literacy. Dutta and Sivaramakrishnan (2013) discussed the changing spatial patterns of the urban-rural gap in literacy concerning the scheduled caste and general caste individuals across the states of India. They show that a high degree of urbanization, greater economic diversification and enlightened administration, greater agricultural prosperity and a higher rate of increase in non-farm employment are vital factors for reducing the urban-rural gap in literacy for both the scheduled caste and non-scheduled caste population. Mitra & Singh (2008) found that poverty is major constraint on the attainment of literacy and leads to a high dropout rate of the scheduled tribe women in the north-eastern states of India. The study reveals that the social and cultural norms and parental negative attitude towards school participation among the tribal women are the chief determinants for not achieving their primary education. Based on the Uttar Pradesh and Bihar survey of living conditions of 1997-98, Dostie & Jayaraman (2006) investigated the determinants of school enrolment in 119 villages. Parental education, household wealth, school quality and infrastructure are seen to enhance enrolments. Some other factors such as parental education, social norms, a lower dependency ratio within the family, employment opportunities, village infrastructure, teachers' regularity and mid-day meals are responsible for increasing school participation for girls (Drèze & Kingdon, 2001; Machin & Vignoles, 2004).

Although some facts regarding the regional inequality in education and its determining factors are addressed in the literature, little attention has been given to the inequality of literacy between urban and rural areas across states in India over time. To capture the disparity in education in urban and rural areas of India, this study uses Sopher's Disparity Index which is hardly explained in earlier papers. In the background, this article constructs the Sopher's Index to analyze the inter-state comparison of the urban-rural literacy gap across 15 Indian states during the period between 1981 and 2011. Secondly, it tries to identify the possible factors which may help to reduce the disparity in urban and rural literacy rates in India.

## 3. The scenario of rural and urban literacy in India

To understand the scenario of India's literacy rate, trends in rural and urban literacy rates and growth rate of literacy during the census year[3] between 1981 and 2011, are discussed in the following tables-

### 3.1 Trends of literacy rate in India

**Table-1** India's aggregate literacy rate and its growth rate during 1981 to 2011

| Census Year | 1981[4] | 1991[5] | 2001 | 2011 | Average growth of literacy rate during 1981 to 2011 |
|---|---|---|---|---|---|
| Rural literacy rate | 36.01 | 44.7 | 58.74 | 68.91 | 19.37 |
| Urban literacy rate | 67.2 | 73.1 | 79.92 | 85 | 7.53 |
| India's aggregate literacy rate | 41.43 | 52.22 | 64.83 | 74.04 | 17.52 |

Source: Author's calculation from different rounds of Indian Census data

Table 1 describes that the illiteracy rate in India in the census year of 1981 was surprisingly high at 58%. There was a significant disparity between urban and rural areas in the country, non-literates being almost 33% in urban areas and 64% in rural areas. In the table-1, the Census of India (2011) further showed a large improvement, the overall literacy rate was 74% - around 85% in urban areas and 69% in rural areas. Though the Indian Census data exhibited a nearly 18 per cent growth rate in literacy at the country's aggregate level, around 8 per cent in urban areas and 19 per cent in rural areas between 1981 and 2011, the problem of inter-state disparities in literacy[6] has not been completely eliminated.

### 3.2 Variations of urban-rural literacy rate in India: An inter-state comparison

---

[3] The study has considered the four census rounds of data from 1981 to 2011 as the latest round of census data in 2021 is yet to publish.
[4] Up to 1981, the census of India used the crude literacy concept to calculate the country's literacy rate and it is defined as the percentage of the literate population as the percentage of the total population aged 0 to old age.
[5] Since 1991, the Indian census has been using the effective literacy rate, defined as the total percentage of literate population of an area at a particular time who can read and write with understanding as the percentage of the total population aged seven years and above.
[6] The literacy rate among the age group seven years and above was recorded at 77.7 per cent at the aggregate level, 73.5 per cent in rural areas, and 87.7 per cent in urban areas (NSSO, July 2017- June 2018).

To analyze the trends of urban and rural literacy rates and their disparity across fifteen Indian states, the study considers two indicators- the ranking of states based on urban and rural literacy rates and the Sopher's disparity index in literacy between urban and rural India.

*3.2.1 Ranking of states based on urban and rural literacy rates*

**Table-2:** Ranking of fifteen States by literacy rates of rural and urban India

| Year / States | 1981 Rural | 1981 Urban | 1991 Rural | 1991 Urban | 2001 Rural | 2001 Urban | 2011 Rural | 2011 Urban |
|---|---|---|---|---|---|---|---|---|
| AP | 13 | 13 | 13 | 13 | 13 | 13 | 15 | 13 |
| Bihar | 14 | 12 | 14 | 12 | 15 | 14 | 14 | 14 |
| Gujrat | 5 | 5 | 5 | 5 | 8 | 6 | 5 | 4 |
| Haryana | 10 | 8 | 8 | 8 | 7 | 10 | 7 | 11 |
| H. P | 2 | 2 | 2 | 2 | 2 | 2 | 2 | 2 |
| Karnataka | 9 | 7 | 9 | 7 | 10 | 8 | 10 | 7 |
| Kerala | 1 | 1 | 1 | 1 | 1 | 1 | 1 | 1 |
| M.P | 11 | 11 | 12 | 11 | 11 | 9 | 12 | 9 |
| Maharashtra | 3 | 3 | 3 | 3 | 3 | 4 | 3 | 3 |
| Odisha | 8 | 10 | 10 | 10 | 9 | 7 | 9 | 6 |
| Punjab | 6 | 9 | 6 | 9 | 5 | 11 | 8 | 10 |
| Rajasthan | 15 | 14 | 15 | 14 | 12 | 12 | 13 | 12 |
| Tamil Nadu | 4 | 4 | 4 | 4 | 4 | 5 | 4 | 5 |
| U.P. | 12 | 15 | 11 | 15 | 14 | 15 | 11 | 15 |
| W.B. | 7 | 6 | 7 | 6 | 6 | 3 | 6 | 8 |

Source: Author's calculation

Table 2 discusses the ranking of the literacy rate of fifteen states of urban and rural India over the census period between 1981 and 2011. As per the census of India (2011), the three top-ranking states in rural and urban educational attainment are Kerala, H.P. and Maharashtra, whereas the three bottom-ranking states are A.P., Bihar, and Rajasthan. Except for H.P. and Kerala, the remaining states have faced substantial variations in ranking positions in rural and urban literacy rates in India over time.

*3.2.2 Trends of variations of inequality in urban-rural literacy rates across the Indian states*

The present study is related to the urban-rural differential in literacy rates in the fifteen major Indian states. The differential rate can be calculated in the absolute or relative sense. In an absolute sense, it may be calculated by subtracting the rural literacy rate ($R_L$) from the urban literacy rate ($U_L$). It can be expressed as ($U_L - R_L$), where, $U_L$ = percentage of literate persons in the urban population, $R_L$ = percentage of literate persons in the rural population.

Similarly, the relative differential in literacy is calculated by using the ratio between the two rates. These two types of formulae which are very commonly used to focus on the differential in literacy are perhaps unrealistic measures. For example, a state with an urban literacy rate of 80 per cent and a rural literacy rate of 60 per cent would have the same absolute differential as another state with an urban literacy of 50 per cent and a rural literacy rate of 30 per cent in a particular year. Similarly, the relative differential in a particular state with 70 (urban) and 35 (rural) per cent literacy rates would be the same as in another state with 40 (urban) and 20 (rural) per cent literacy rates in a year.

Hence, a more appropriate technique is required to measure the inequality in educational attainment. To analyze the inter-state comparison of urban-rural disparity in educational attainment in India, the study has calculated the Sopher's index. This is the logarithm[7] (of base 10) of the odds ratio which measures inequality between two different groups of the population (Sopher, 1974). It can be expressed as-

$$DLI_{U-R} = \log\left(\frac{U_L}{R_L}\right) + \log\left(\frac{100-U_L}{100-R_L}\right)$$

Where; $DLI_{U-R}$ = Sopher's urban-rural differential literacy index

$U_L$ = Percentage of literates in urban population

$R_L$ = Percentage of literates in rural population

Where, $U_L \geq R_L$

In this method, to measure the inequality in education, we consider urban literacy rates of all selected Indian states to have comparatively higher values than the rural literacy rates since we have already found (in Table-1, Table-6 and Table-7 in Appendix) in our study that the urban literacy rate is higher than the rural literacy rate.

---

[7] The main idea behind taking log is to lessen the levelling-off effect, i.e. the states with higher literacy rates may show a lower level of inequality than the states with low literacy rates even though the gap is the same for both states (Sopher, 1980).

In case of perfect equality in education between rural and urban regions of all states in India, the value of $DLI_{U-R}$ will be zero. The higher value of $DLI_{U-R}$ indicates the high urban-rural disparity in educational attainment and the lower the value, the lower is the educational disparity across the states in India. Usually, this method is suitable in measuring the relative inequality. For example, if urban and rural literacy disparity is calculated only for the year 1981, the value of $DLI_{U-R}$ might not give much meaningful interpretation unless we compare it with the urban and rural inequality in literacy for another year (such as 1991). While comparing the disparity indices in literacy, it is crucial to understand that in the year for which $DLI_{U-R}$ is less, the disparity is comparatively low and the year in which $DLI_{U-R}$ is observed more, the disparity is showing high.

**Table 3:** State-level urban-rural disparity as measured by Sopher's index of educational attainment in the fifteen states of India in different census years

| States | 1981 | 1991 | 2001 | 2011 |
|---|---|---|---|---|
| AP | 0.606 | 0.550 | 0.424 | 0.420 |
| Bihar | 0.615 | 0.617 | 0.515 | 0.361 |
| Gujarat | 0.553 | 0.460 | 0.454 | 0.417 |
| Haryana | 0.566 | 0.449 | 0.345 | 0.267 |
| H Pradesh | 0.590 | 0.516 | 0.427 | 0.341 |
| Karnataka | 0.564 | 0.499 | 0.454 | 0.436 |
| Kerala | 0.413 | 0.171 | 0.180 | 0.162 |
| Madhya Pradesh | 0.623 | 0.637 | 0.449 | 0.449 |
| Maharashtra | 0.573 | 0.484 | 0.394 | 0.417 |
| Odisha | 0.542 | 0.489 | 0.452 | 0.422 |
| Punjab | 0.508 | 0.364 | 0.315 | 0.274 |
| Rajasthan | 0.647 | 0.636 | 0.412 | 0.403 |
| Tamil Nadu | 0.566 | 0.469 | 0.382 | 0.377 |
| Uttar Pradesh | 0.545 | 0.432 | 0.319 | 0.205 |
| W.B. | 0.580 | 0.475 | 0.554 | 0.315 |
| India | 0.582 | 0.527 | 0.446 | 0.408 |

Source: Author's calculation

Table 3 reflects that among all the states, Madhya Pradesh, Karnataka and Andhra Pradesh, are examples of high indexes of urban-rural differential in literacy in the period between 1981 and 2011. The coefficient of inequality in literacy between urban and rural Madhya Pradesh has decreased from 0.623 to 0.449. There has been a decrease in the urban-rural differential index from 0.564 to 0.436 in Karnataka from the year 1981 to 2011. A similar trend is also observed in Andhra Pradesh; the urban-rural literacy gap has decreased from 0.606 in the 1981 census to 0.420 in the 2011 census. A significant decrease in the urban-rural literacy gap has been found in states such as Kerala (from 0.413 to 0.162), U.P (from 0.545 to 0.205), Haryana (from 0.566 to 0.267), Punjab (from 0.508 to 0.274), and W.B (from 0.580 to 0.315) in the period between 1981 and 2011.

## 4. Data and Methodology

The rural and urban areas of 15 major states are considered in the study. To analyze the inequality in educational attainment in India over time, the required dataset and their sources are described as follows-

i. Data related to the state literacy rate, and the rural and urban literacy rate of the four Indian census years of 1981, 1991, 2001 and 2011 are taken from the Office of the General and Census Commissioner, India.

ii. Data for the total fertility rate of urban and rural areas in the years 1981, 1991, 1999 and 2009 are drawn from the Ministry of Health and Family Welfare, Govt. of India. By interpolation method, data for 2001 and 2011 are hypothesized from the data in 1999 and 2009.

iii. Data related to the Per Capita Net State Domestic Product (PCNSDP) at factor cost by the years 1980-81, 1990-91, 2000-01 and 2010-11 are drawn from the Reserve Bank of India.

iv. Percentages of married females by age are divided into the three age groups- below 18 years, 18-20 years and 21+ years, taken from the Sample Registration System, Office of the Registrar General of India in 1993, 2000. From the datasets, the required data for the years 1981, 1991, 2001 and 2011 are postulated by applying the interpolation technique.

v. Teacher-pupil ratio data for 1982-83, 1990-91, 2000-01 and 2011-12 are drawn from the Ministry of Human Development, Govt. of India. These data are used as the proxy data for 1981, 1991, 2001, and 2011 respectively.

vi. Male and female labour force participation rate is taken from the three rounds of the National Sample Survey (NSS) of 1993-94 (50th), 1999-00 (55th) and 2011-12 (68th), used as the proxy data of 1991, 2001 and 2011. Based on the three NSSO survey points, data for the year 1981 is estimated using the interpolation technique.

vii. Data related to state-level social sector expenditure in 1990-91, 2000-01 and 2010-11 are taken from the 'Handbook of Statistics on State Government Finances-2010 and various issues of 'State Finances: A Study of Budgets', Reserve Bank of India. Based on these data, 1981 data is estimated using the interpolation method.

Now, it is required to identify the responsible factors that may reduce the disparity in the urban-rural literacy rate in India. The theoretical explanations of the regressors are as given below-

a) Per capita Net State Domestic Product of the $i^{th}$ states ($PCNSDP_i$): It is measured as the ratio of Net State Domestic Product and the total population in a particular state in a particular year. Per capita NSDP at factor cost with the base years of 1980-81, 1993-94, 2000-01 and 2004-05 are converted into a constant price with 1993-94 as the base year. Here, the per capita NSDP is taken as the proxy of family income[8]. It is observed that low-income household parents hardly spend quality time on their children's educational activities and allocate fewer resources to education (Banerji & et al., 2013).

b) Percentages of married females by age of $i^{th}$ states ($MFA_i$): In the paper, we have considered only number of rural females who got married only at the age of 21 years or after the age of 21 years ($MFA_{21+_i}$) in fifteen states of India. According to Jensen and Thornton (2003), later ages of marriage are positively associated with women's higher education. They observed that women whose marriage is delayed until after age 21 receive three to four years more education than an Indian girl who marries before the age of 15. Hence, later marriage is likely to be a significant determinant to promote the female's education, especially in rural areas.

c) The total fertility rate (or children per rural woman of $i^{th}$ states) ($TFR_i$): TFR is the average number of children that women would bear if they experience the current age-specific fertility rates and survive to the end of their reproductive life. Although the fertility rate for the country as a whole has declined over the period, the rural fertility rate[9] is higher than the urban fertility rate. The high fertility rate is expected to be negatively associated with education as a high number of children may not equally access education facilities, especially in poor households (Gotmark & Andersson, 2020).

---

[8] Here it is conceded that there might be inequality of household incomes within a state.

d) Literacy rate for rural females of $i^{th}$ states ($Flit_i$): It measures the females who at least read and write with the understanding of a language as a percentage of total rural females. The literacy rate for rural women is used here as the proxy for a mother's educational attainment in a particular state in a particular year. A mother's education is more strongly associated with the child's health and learning outcome than the father's education (Breierova & Duflo, 2004; Banerji et al., 2013).

e) Teacher-pupil ratio of $i^{th}$ states ($TPratio_i$): It is a function of the number of students per teacher. Teacher-pupil ratio is an important indicator of teachers' effectiveness and students' achievements. Koc & Celik (2015) observed that there is a negative correlation between the student-teacher ratio and students' achievement with a large number of students per teacher resulting in low achievement in a school. A lower student-teacher ratio is expected to optimize learning outcomes in the class that may motivate parents to send their children to school.

f) Rural male labour force participation of $i^{th}$ states of India ($Mlfpr_i$): It defines the ratio of the total male labour force participation in farm and non-farm sector to the total rural male labour force. It is used as a proxy of the father's monthly income, and it may positively influence children's educational outcomes through the income effect (Afridi et al., 2013).

g) Rural female labour force participation of $i^{th}$ states of India ($Flfpr_i$): It is calculated as the total female labour force participation in farm and non-farm sectors is divided by the total working-age female population in rural India. The female labour force is used as a proxy of a mother's employment. Afridi & et al. (2013) reported that mother's labour force participation negatively impacts children's education if children, especially, girl children are engaged in domestic chores when mothers are at the workplace. Hence, it is required to examine whether rural female workings for earnings

---

[9] The TFR in rural areas is 2.7 whereas 1.9 in urban India (Office of the Registrar General & Census Commissioner, India, 2011).

outside the home impact a child's educational attainment.

h) Social sector expenditure of $i^{th}$ states (Socialexp$_i$): The country's development comes through social sector expenditure incurred on civic amenities, labour welfare, and human resource development (Kundu & Das, 2019). Kaur & Misra (2003) show that public spending on social sector expenditure on school infrastructure is more productive in poorer states than in non-poorer states. It is observed that the share of govt. expenditure on the educational sector in GDP is only 3.1 per cent (Economic Survey, 2019-20) against the recommendation of 6 per cent by the National Policy on Education in 1968. Underinvestment in education is one of the obstacles for not achieving the educational goals and targets of universal adult literacy and Universal Elementary Education (Tilak, 2006). More than 10 lakh government schools where almost 24.8 crore enrolled children are underspent in school development (Unified District Information System for Education [DISE], 2020-21). Hence, the paper seeks to examine whether the inadequacy of social sector investment in education is responsible for the continuation of adult illiteracy, high dropout rates of children, and inter-state disparity in educational attainment.

Now summary statistics of regressors as well as dependent variables are presented in Table 4.

**Table 4**: Summary Statistics of the explanatory variables and dependent variable- urban-rural differential literacy index (DLI$_i$).

| Year | 1981 | | | | 1991 | | | |
|---|---|---|---|---|---|---|---|---|
| Variables | Max | Min | Mean | S. D | Max | Min | Mean | S. D |
| DLI | 0.75 | 0.291 | 0.5407 | 0.169 | 0.73 | 0.271 | 0.52 | 0.169 |
| PCNSDP (Rs) | 3119 | 1044 | 1872.9 | 587.8 | 9872 | 2868 | 5722.8 | 1926.1 |
| MFA$_{21+}$ (%) | 81.1 | 21.9 | 39.52 | 15.045 | 64.4 | 5.2 | 22.83 | 15.05 |
| TFR (%) | 6 | 2.9 | 4.58 | 0.879 | 5.4 | 1.8 | 3.65 | 0.98 |
| Flit (%) | 73.36 | 13.36 | 31.16 | 14.58 | 86.17 | 20.44 | 42.48 | 16.49 |
| TPratio | 50 | 32 | 39.125 | 4.425 | 53 | 32 | 42.44 | 5.78 |
| Mlfpr (%) | 75.03 | 29.13 | 54.103 | 16.98 | 73 | 27.1 | 52.07 | 16.98 |
| Flfpr (%) | 90.97 | 80.87 | 87.77 | 2.87 | 90.3 | 80.2 | 87.106 | 2.875 |
| Socialexp (Billion) | 22.87 | 0.17 | 9.37 | 6.99 | 47.1 | 4.4 | 21.66 | 11.66 |
| Year | 2001 | | | | 2011 | | | |

| Variables | Max | Min | Mean | S. D | Max | Min | Mean | S. D |
|---|---|---|---|---|---|---|---|---|
| DLI | 0.675 | 0.244 | 0.488 | 0.15 | 0.64 | 0.12 | 0.41 | 0.15 |
| PCNSDP (Rs) | 28943 | 6200 | 18444 | 6630.1 | 106358 | 22582 | 65547.9 | 24825.6 |
| MFA$_{21+}$(%) | 68.6 | 13.1 | 31.56 | 16.57 | 71.23 | 15.73 | 34.19 | 16.57 |
| TFR (%) | 4.6 | 1.9 | 3.13 | 0.874 | 3.8 | 1.8 | 2.48 | 0.739 |
| Flit (%) | 87.86 | 33.57 | 57.16 | 12.87 | 92.07 | 51.5 | 67.03 | 10.53 |
| TPratio | 67 | 25 | 42.63 | 11.27 | 59 | 16 | 29.69 | 10.18 |
| Mlfpr (%) | 67.5 | 24.4 | 48.82 | 15.045 | 64.3 | 10.8 | 40.76 | 14.89 |
| Flfpr (%) | 88.5 | 78 | 84.96 | 2.92 | 87 | 75.5 | 82.23 | 3.36 |
| Socialexp (Billion) | 154.3 | 19.6 | 76.23 | 36.046 | 538.3 | 60.2 | 260.29 | 143.7 |

Source: Author's calculation

The study considers fifteen Indian states during four census years from 1981 to 2011. For the investigation of possible determinants of urban-rural educational inequality, Fixed effect panel data regression analysis has been applied here, and the equation can be written as-

$$DLI_{it} = \beta_1 x'_{it} + \omega_i + \varepsilon_{it}, \text{ and } i = 1, 2, 3…,15; t = 1, 2, 3, 4 \quad …………..… (1)$$

Where, $DLI_{it}$ indicates the differential literacy index for i$^{th}$ states in the t$^{th}$ time period, $x'_{it}$ is the time-variant regressor vector, $\omega_i$ captures the unobserved heterogeneity which is fixed over time but varies cross-sectionally and $\varepsilon_{it}$ represents as idiosyncratic error term, where, $\varepsilon_{it} \sim i.i.d., N(0.\sigma_e^2)$, and exogeneity of the covariates $x_{it}$, i.e., $cov(\varepsilon_{it}, x_{it}) = 0$.

Initially, the multicollinearity test has examined whether there are any inter-correlations among the explanatory variables. The Variance Inflation Factor [VIF = $\frac{1}{(1-R^2)}$] shows that two variables such as literacy rate of rural females (Flit$_i$) and total fertility rate (TFR$_i$) are suffering from multicollinearity problems as their VIF values are more than 4.

To reduce the biasedness, two separate Fixed Effect[10] panel data regression models are considered in the study. Now three equations can be written as-

$$DLI_{1it} = \beta_0 + \beta_1 PCNSDP_{it} + \beta_2 MFA_{21+_{it}} + \beta_3 TFR_{it} + \beta_4 TPratio_{it} + \beta_5 Mlfpr_{it}$$

---

[10] Hausman test confirms that the fixed effect panel data regression analysis is more appropriate over the random effect panel data technique.

$$+ \beta_6 \text{Flfpr}_{it} + \beta_7 \text{Socialexp}_{it} + \varepsilon_{it} \ldots \ldots \ldots \ldots \ldots \ldots (2)$$

$$\text{DLI}_{2it} = \beta'_0 + \beta'_1 \text{PCNSDP}_{it} + \beta'_2 \text{MFA}_{21+_{it}} + \beta'_3 \text{Flit}_{it} + \beta'_4 \text{TPratio}_{it}$$

$$+ \beta'_5 \text{Mlfpr}_{it} + \beta'_6 \text{Flfpr}_{it} + \beta'_7 \text{Socialexp}_{it} + \varepsilon'_{it} \ldots \ldots \ldots (3)$$

The result of the Fixed effect panel data regression model has been shown in the following table 5:

**Table 5:** Result of Fixed Effect panel data regression analysis: Dependent variable- Urban-Rural differential literacy index for $i^{th}$ states in India

| Models | Equation (2) | Equation (3) |
|---|---|---|
| Observations | 60 | 60 |
| Variables | Value of the Coefficient | Value of the Coefficient |
| PCNSDP | 7.94e$^{-07}$ (4.87e$^{-07}$) | 5.13e$^{-07}$ (5.82e$^{-07}$) |
| MFA$_{21+}$ | -0.00069* (0.00035) | -0.0048*** (0.0009) |
| TFR | - | 0.0776* (0.0205) |
| Flit | -0.0073*** (0.0012) | - |
| TPratio | 0.0014 (0.0012) | -0.0003 (0.0012) |
| Mlfpr | 0.0007 (0.0067) | 0.0101 (0.0078) |
| Flfpr | -0.0026* (0.0019) | -0.0008* (0.0003) |
| Socialexp | -0.00006 (0.00012) | -0.00003 (0.0001) |
| Constant | 0.6071* | 0.7257* |
| R-Square (within) | 0.8575 | 0.7925 |
| F (14, 38) | 3.83*** | 3.42*** |

**Note**: *** indicates 1% level of significance, ** indicates 5% level of significance and * indicates 10% level of significance.

## 5. Discussion of results

The results of the fixed effect panel data regression analysis show that the percentage of females who married after the age of 21 (MFA$_{21+}$) is observed to have significant and negatively related to the urban-rural literacy gap. This relation confirms that later ages of marriage are strongly associated with women's greater levels of school education which significantly affects the reduction of the urban-rural literacy gap. The total fertility rate for rural females (TFR) is found to be positively significant in determining urban-rural differential in education (Martin, 1995). Rural female labour force participation used as the proxy of mothers' employment is positively associated with the children's educational attainment as it is expected that the employed mothers are more interested to invest money for their children's better education. Female literacy in rural areas represents mothers' literacy is another key factor to increase the literacy rate in rural areas as they are more aware to educate their children. The Annual Status of Rural Education Report (2019) stated that the mother's educational attainment is sometimes responsible to determine the type of pre-schooling or schooling that the child receives. This report further illustrated that mothers who completed eight or less than eight years of schooling are willing to attend their children to anganwadis or government pre-primary classes, whereas, the mothers who studied secondary or higher secondary education are likely to enrol them in private lower/upper kindergarten classes to provide quality education.

## 6. Conclusion and policy implications

This study aims to examine the nature of the variations of urban-rural differential in literacy across fifteen major states in India and investigates its responsible factors for reducing the urban and rural disparity in education. The Sopher's index has been constructed to analyze the changes in inequality of educational attainment across fifteen states in India over time. The states such as Rajasthan, Madhya Pradesh, Maharashtra and Gujarat have faced a high urban-rural differential in literacy, whereas, a significant improvement in rural literacy has been found in U.P, Haryana, Bihar, and Punjab in the period between 1981 and 2011. The fixed effect panel data regression technique has been applied here to illustrate the determinants that may positively influence the reduction of the urban and rural literacy gap in India. The findings of the model suggest four policy recommendations that boost rural literacy. The lower fertility rate of the rural females represents the smaller number of children in the family, lower expenditure burden of households and higher opportunity for their children to attend school. Marriage after the age of 21 among rural females and female literacy helps to reduce urban and rural education disparity as these two factors empower females in the decision-making process in the households and make them aware of their children's education. Employment opportunities among rural females (used as a proxy of the mother's job) also significantly enhance household income, thereby encouraging them to continue their children's education and helping to bridge the literacy gap between those living in rural and urban areas.

**Appendix**

**Table 6**: State wise Rural and Urban Literacy rates in 1981 and 1991

| Census years | 1981 | | 1991 | |
|---|---|---|---|---|
| States | Rural literacy rate | Urban literary rate | Rural literacy rate | Urban literacy rate |
| AP | 27.85 | 61 | 35.74 | 66.35 |
| Bihar | 27.7 | 61.77 | 33.83 | 67.89 |
| Gujarat | 43.57 | 71 | 53.09 | 76.54 |
| Haryana | 37.26 | 66.83 | 49.85 | 73.66 |
| H Pradesh | 48.89 | 77.8 | 61.86 | 84.17 |
| Karnataka | 37.63 | 66.91 | 47.69 | 74.2 |
| Kerala | 80.31 | 86.91 | 88.92 | 92.25 |
| M. P | 29.33 | 64.55 | 35.87 | 70.81 |
| MH | 45.65 | 74.29 | 55.52 | 79.2 |
| Orissa | 37.77 | 64.81 | 45.46 | 71.99 |
| Punjab | 41.73 | 64.96 | 52.77 | 72.08 |
| Rajasthan | 22.47 | 58.05 | 30.37 | 65.33 |
| T. N | 45 | 73.25 | 54.59 | 77.99 |
| U. P | 28.53 | 54.87 | 36.66 | 61 |
| W. B | 40.18 | 70.68 | 50.5 | 75.27 |
| All India | 36.01 | 67.2 | 44.7 | 73.1 |

**Table 7**: State wise Rural and Urban Literacy rates in 2001 and 2011

| Census years | 2001 | | 2011 | |
|---|---|---|---|---|
| States | Rural literacy rate | Urban literary rate | Rural literacy rate | Urban literacy rate |
| AP | 54.5 | 76.09 | 61.14 | 80.54 |
| Bihar | 43.92 | 71.93 | 61.8 | 78.8 |
| Gujarat | 61.29 | 81.84 | 73 | 87.6 |
| Haryana | 63.19 | 79.16 | 72.74 | 83.14 |
| H Pradesh | 75.08 | 88.95 | 82.91 | 91.4 |
| Karnataka | 59.33 | 80.58 | 68.86 | 85.78 |
| Kerala | 90.04 | 93.19 | 92.9 | 95 |
| M. P | 57.8 | 79.39 | 65.3 | 84.1 |
| MH | 70.36 | 85.48 | 77.1 | 89.8 |
| Orissa | 59.84 | 80.84 | 70.8 | 86.5 |
| Punjab | 64.72 | 79.1 | 72.45 | 83.18 |
| Rajasthan | 55.34 | 76.2 | 62.3 | 80.7 |
| T. N | 66.21 | 82.53 | 73.8 | 87.04 |
| U. P | 52.53 | 69.75 | 67.6 | 77 |
| W. B | 63.42 | 86.13 | 72.97 | 84.78 |
| All India | 58.74 | 79.92 | 68.91 | 85 |